\documentclass[debug,overfull]{epl}
\usepackage{amsmath}
\usepackage{amssymb}

\title{Critical level spacing distribution \\
in long-range hopping Hamiltonians}
\shorttitle{Critical level spacing distribution}
\author{E. Cuevas}
\institute{
Departamento de F{\'\i}sica, Universidad de Murcia -
E-30071 Murcia, Spain.}
\begin{document}

\maketitle

\begin{abstract}
The nearest level spacing distribution $P_c(s)$ of $d$-dimensional
disordered models ($d=1$ and $2$) with long-range random hopping
amplitudes is investigated numerically at criticality. We focus on
both the weak ($b^d \gg 1$) and the strong ($b^d \ll 1$) coupling
regime, where the parameter $b^{-d}$ plays the role of the coupling
constant of the model. It is found that $P_c(s)$ has the asymptotic
form $P_c(s)\sim\exp [-A_ds^{\alpha}]$ for $s\gg 1$, with the critical
exponent $\alpha=2-a_d/b^d$ in the weak coupling limit and
$\alpha=1+c_d b^d$ in the case of strong coupling.
\end{abstract}

It is well established that the statistical properties of spectra
of disordered one-electron systems are closely related to the
localization properties of the corresponding wavefunctions
\cite{AS86,SS93,KL94}. In the metallic phase, the large overlap of
delocalized states, which are essentially structureless, induces
correlations in the spectrum, leading to the well known level
repulsion effect. If the system is invariant under rotation and
under time-reversal symmetry (orthogonal symmetry), the normalized
spacings $s$ follow Wigner-Dyson statistics at the infinite system
size limit:
\begin{equation}
P_{\rm W}(s) = \frac{\pi}{2}s\exp \left [-\frac{\pi}{4}s^2 \right ]\;.
\label{wigner}
\end{equation}
In contrast, in the localized regime, states with close energy levels
are typically localized at different parts of space and have an
exponentially small overlap. Their levels are therefore uncorrelated
and the corresponding spacings are distributed according to the
Poisson law
\begin{equation}
P_{\rm P}(s) = \exp[-s]\;.\label{poisson}
\end{equation}

It has been argued that the statistics of energy levels at the 
disorder-induced  metal-insulator transition (MIT) is characterized
by a third universal (i.e., independent of the system size and of
the details of the Hamiltonian model) distribution $P_c(s)$, which
is different from both Wigner-Dyson statistics and the Poisson
statistics \cite{AZ88,SS93}. The asymptotic behavior of this
distribution for $s \gg 1$ has been a controversial issue and 
still remains unresolved. On the one hand, the influence of the MIT
on the spectral properties was studied in Refs. \cite{AZ88,SS93} by
means of the impurity diagram technique combined with scaling
assumptions. In these studies, it was conjectured that
\begin{equation}
P_c(s)\sim \exp [-\kappa s]\;, \quad s \gg 1 \;,
\label{tailsh}
\end{equation}
with $\kappa \approx 3.3$, the reason for such behavior being that
the Thouless energy at the transition point is of the order
of the average level spacing ($E_c/\Delta \approx 1$), and so the
levels's repulsion is effective only for $s \lesssim 1$.

On the other hand, by mapping the energy level distribution onto the
Gibbs distribution for a classical one-dimensional gas with a repulsive
pairwise interaction, ref. \cite{AK94} derived the following asymptotic
form for $P_c(s)$:
\begin{equation}
P_c(s)\sim \exp [-A_ds^{\alpha}]\;, \quad s \gg 1\;,
\label{tailar}
\end{equation}
where the coefficient $A_d$ depends only on the dimensionality, $d$,
and where the critical exponent, $\alpha$, which ranges in the interval
$1<\alpha<2$, is related to the correlation length exponent $\nu$
and to the dimensionality through $\alpha=1+(d\nu)^{-1}$.

As regards the numerical description of $P_c(s)$, there is also no
consensus. The exponential decay,
eq. (\ref{tailsh}), of $P_c(s)$ has been confirmed by most
groups at different MITs (see Ref. \cite{Ni99} and references therein),
while an exponent $\alpha \approx 1.2$ has been found in refs.
\cite{Ev94,VH95} from a fit in the whole range of spacings to a
distribution of the form $P_c(s)=Bs\exp[-As^{\alpha}]$ or, indirectly,
from the two-point correlation function of the density of states \cite{BM95}.
Anyway, the behavior (\ref{tailar}) with some nontrivial
$1 \le \alpha \le 2$ is what one would expect at the mobility edge.

It should be pointed out that MITs generically take place at strong disorder
(conventional Anderson transition, quantum Hall transition, transition
in $d=2$ for electrons with strong spin-orbit coupling, etc.). In this
regime, the predicted \cite{AK94} exponent $\alpha=1+(d\nu)^{-1}$ slightly
deviates from unity, making it relatively difficult to see on the
numerically calculated tails of $P_c(s)$ ({\it e.g.}, at the standard Anderson
transition in 3D $\alpha \approx 1.2$). To overcome this problem, it is
necessary to investigate transitions which occur at the opposite limit
(weak coupling regime). This area has been left unexplored, but one would
expect to find an exponent $\alpha$ far from unity and closer to the
Wigner-Dyson value $\alpha=2$. 

In this work, we try to definitively solve the existing controversy about
the large $s$ asymptotic form of $P_c(s)$. From results of detailed
high-precision numerical investigations, we will show unambiguously that
eq. (\ref{tailar}) is indeed  correct, while the validity of eq. (\ref{tailsh})
is limited to the case of very strong disorder (strictly at the limit of
infinity coupling strength). In addition, we find that the exponent $\alpha$
in eq. (\ref{tailar}) continuously varies between 1 and 2 as the coupling
strength of the Hamiltonian model changes from 0 to $\infty$.

To this end, we performed numerical calculations of $P_c(s)$ on a
generalization to $d$ dimensions of the power law random banded matrix
(PRBM) model \cite{MF96,KM97,Mi00,ME00,KT00,Va02,CG01,CO02,Cu03,Cu03a}
(for closely related models see also Ref. \cite{PS94}).
The corresponding Hamiltonian, which describes non-interacting
electrons on a disordered $d$-dimensional square lattice with
random long-range hopping, is represented by real symmetric matrices,
whose entries are randomly drawn from a normal distribution with zero
mean, $\left\langle {\cal H}_{ij} \right\rangle =0$, and a variance
which depends on the distance between the lattice sites $\bm{r}_i$:
\begin{equation}
\left\langle |{\cal H}_{ij}|^2\right\rangle =\frac{1}
                     {1+(|\bm{r}_i-\bm{r}_j|/b)^{2d}}
\times\left\{\begin{array}{ll}
                    \dfrac {1}{2}     \ ,\quad & i\neq j\,,\\
                    1 \,\ ,\quad & i=j \,.
       \end{array}\right.
\label{h1dor}
\end{equation}
We refer the reader to Ref. \cite{Cu03a} (and references therein)
for the advantages of the  present model with respect to Hamiltonians
with short-range, off-diagonal matrix elements, and for the many real
systems of interest that can be described by Hamiltonians (\ref{h1dor}).

The parameter $b^d$ in eq. (\ref{h1dor}) is an effective bandwidth that
serves as a continuous control parameter over a whole line of criticality,
{\it i.e.}, for an exponent equal to $d$ in the hopping elements
${\cal H}_{ij} \sim b^d$ \cite{Le89}. Furthermore, it determines the critical
dimensionless conductance in the same way as the dimensionality labels the
different Anderson transitions. Each regime is characterized by its
respective coupling strength, which depends on the ratio
$(\langle |{\cal H}_{ii}|^2 \rangle/\langle |{\cal H}_{ij}|^2 \rangle)^{1/2}
\propto b^{-d}$ between diagonal disorder and the off-diagonal transition
matrix elements of the Hamiltonian \cite{Ef83}.

We remind the reader that in the two limiting cases of the 1D model, $b \gg 1$
and $b \ll 1$, which correspond to the weak- and the strong-disorder limits,
respectively, some critical properties (spectral compressibility, correlation
dimension, ...) have been derived analytically by mapping Hamiltonian
(\ref{h1dor}) onto an effective $\sigma$-model of a one-dimensional nature
\cite{Mi00,ME00,KM97,MF96,KT00}. We stress that, unlike the 1D PRBM model,
it has not until now been possible to analytically solve the 2D disordered
models with long-range transfer terms.

The system size ranges between $L=1000$ and $4000$ in 1D, and between
20 and 100 in 2D, whereas  $b^d$ ranges the interval
$0.02 \le b^d \le 10$. We consider a small energy window, containing
about 10\% of the states around the center of the spectral band. The number
of random realizations is such that the number of critical levels included
for each $L$ is roughly $1.2\times 10^6$, except for the larger system size
in 2D, for which this number is about $3\times 10^5$. In order to reduce edge
effects, periodic boundary conditions are included.

For the computation of $P_c(s)$, we unfold the spectrum in each case to
a constant density, and rescale it so as to have the mean spacing equal
to unity. In order to diminish the magnitude of the relative fluctuations
and to analyze the asymptotic behavior in detail, it is more convenient to
consider the cumulative level spacing distribution function
$I(s)=\int_{s}^{\infty} P(s')ds'$. Note that the integration does not change
the asymptotic behavior of $P(s)$. The Wigner surmise, eq. (\ref{wigner}),
and the Poisson distribution, eq. (\ref{poisson}), yield
$I_{\rm W} (s)=\exp [-\pi s^2/4]$ and $I_{\rm P}(s)=\exp [-s]$, respectively.

\begin{figure}[t]
\onefigure[width=7cm,height=7cm,angle=0]{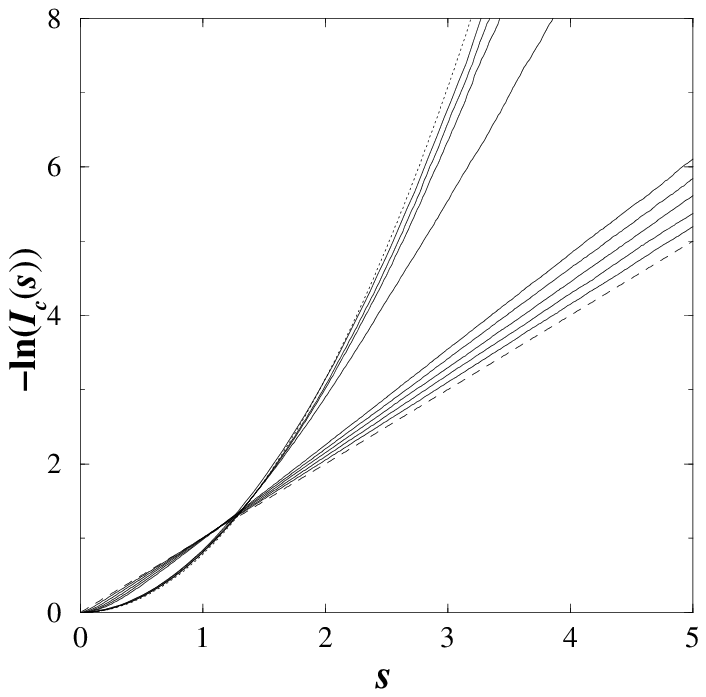}
\caption{The integrated probability $I_c(s)$ of the 2D system for $L=60$ at
$b^2=0.02$, 0.04, 0.06, 0.08, 0.1, 0.4, 0.8, 1 and 10 (from bottom
to top). Dotted and dashed lines are $I_{\rm W}(s)$ and $I_{\rm P}(s)$,
respectively.}
\label{fig1}
\end{figure}

Figure \ref{fig1} displays our results for the integrated probability
$I_c(s)$ of the 2D system for $L=60$ at $b^2=0.02$, 0.04, 0.06, 0.08,
0.1, 0.4, 0.8, 1 and 10, which are depicted consecutively from bottom
to top. Dotted and dashed lines, which correspond to $I_{\rm W}(s)$ and
$I_{\rm P}(s)$, respectively, are given for comparison. A gradual crossover
in the large $s$ tail of $I_{\rm c}(s)$ from the Poisson to the Wigner-Dyson
limiting forms as one increases the inverse coupling constant $b^2$ of the
model can clearly be seen. So, we can therefore expect an exponent $\alpha$
in eq. (\ref{tailar}), which spans the interval $[1,2]$, in agreement
with ref. \cite{AK94}.

\begin{figure}[t]
\twofigures[width=7cm,height=7cm,angle=0]{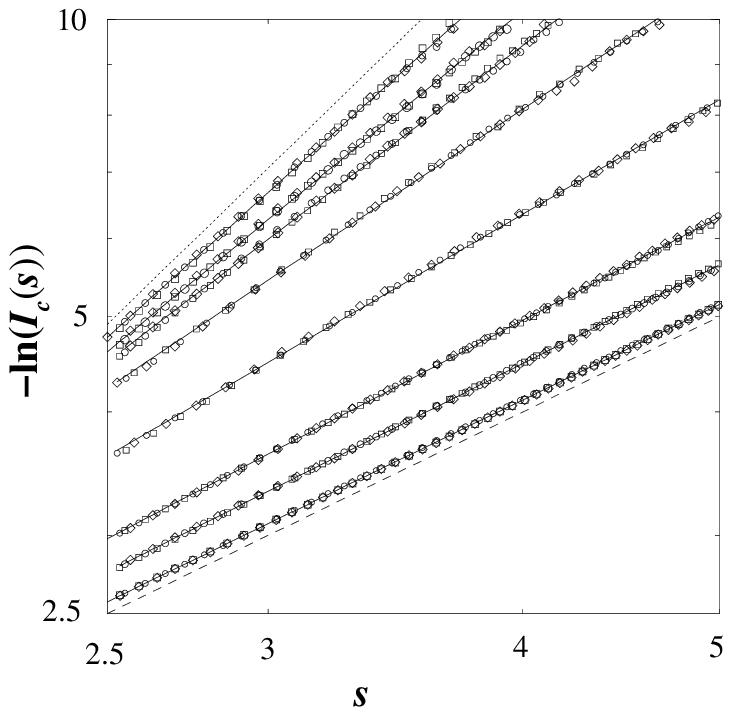}{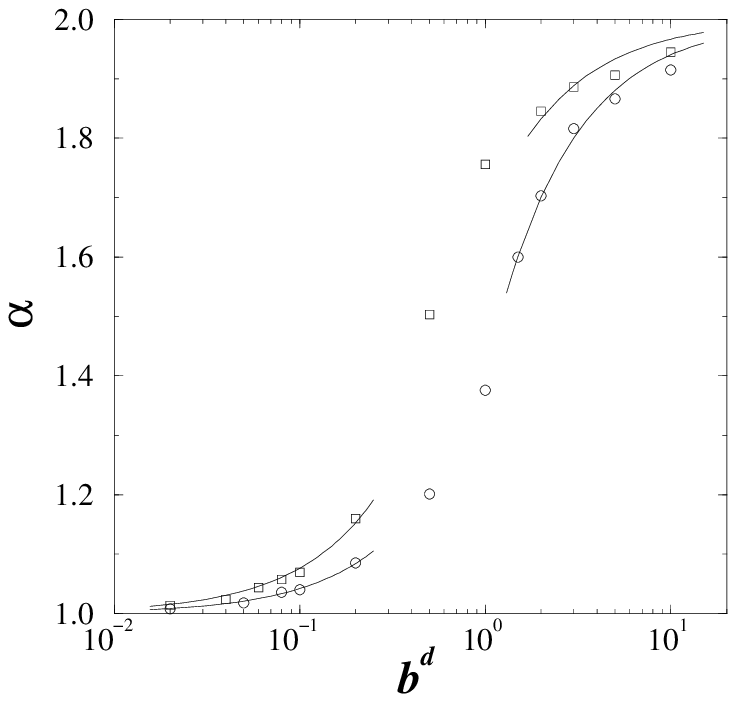}
\caption{Log-log plot of the integrated probability $I_c(s)$ of the 1D system
at $b=0.02$, 0.1, 0.2, 0.5, 1.0, 1.5, 2 and 5 (from bottom to top) and different
system sizes $L=1000$ (circles), 2000 (squares) and 4000 (diamonds). Dotted and
dashed lines are $I_{\rm W}(s)$ and $I_{\rm P}(s)$, respectively, and the straight
lines are fits to eq. (\ref{tailar}).}
\label{fig2}
\caption{The $b^d$-dependence of the critical exponent
$\alpha$ for the 1D (circles) and 2D (squares) disordered systems.
Solid lines are fits to eqs. (\ref{alfavbd}) corresponding to the
limiting cases of weak ($b^d \gg 1$) and strong ($b^d \ll 1$) disorder.}
\label{fig3}
\end{figure}

Next we consider the behavior of $I_c (s)$ with system size $L$.
The results for $s$ large of the critical $I_c (s)$ for the 1D system
at different values of $b$ are shown in a log-log scale in
fig. \ref{fig2} for different system sizes: $L=1000$ (circles), 2000
(squares) and 4000 (diamonds). As in the 2D case (see fig. \ref{fig1}),
one can viasualize the crossover between the small-$b$ and large-$b$
asymptotics. Note that $I_c(s)$ is an $L$-independent
universal scale-invariant function that interpolates, as previously
mentioned, between Wigner and Poisson limits . This result confirms the
existence of a critical distribution exactly at the transition. Dotted
and dashed lines correspond to $I_{\rm W}(s)$ and $I_{\rm P}(s)$,
respectively. We checked that the normalized variances of $P_c(s)$ are
indeed scale-invariant at each critical point studied \cite{Cu99}.
The straight line behavior of the data in such
a plot at all values of $b$ considered is undoubtedly consistent
with a $b$-dependent exponent $\alpha$ in eq. (\ref{tailar}). The values
of $b$ reported are $0.02$, $0.1$, $0.2$, $0.5$, $1.0$, $1.5$, $2$ and $5$,
from bottom to top. The best fit to eq. (\ref{tailar}) in the interval
$2.5 \lesssim s \lesssim 5$ for small $b$ and $2.5 \lesssim s \lesssim 4$
for large $b$, yields $\alpha=1.008$, 1.040, 1.085, 1.201, 1.376, 1.600,
1.703 and 1.866, respectively, thus confirming the result of \cite{AK94}.
Note that for the large energy ranges considered, where $I_c(s)$ vary by
one to three orders of magnitude, the quality of the fits, which are
represented as solid straight lines, is evident.

The disorder dependence of the critical exponent, $\alpha$, as obtained
from the previous fits for the 1D (circles) and 2D (squares) systems
is shown in fig. \ref{fig3} in the broad range of the parameter $b^d$
of the PRBM model. For both dimensions, $d=1$ and 2, it clearly
changes continuously from 1 as $b^d \to 0$ to 2 as $b^d \to \infty$.
In the two limiting cases of weak ($b^d \gg 1$) and strong ($b^d \ll 1$)
disorder regimes it can be well fitted by
\begin{equation}
\alpha=\left\{\begin{array}{ll}
                     2-a_d/b^d \ ,\quad & b^d \gg 1 \,,\\
                     1+c_d b^d \;\;\; ,\quad & b^d \ll 1 \,,
       \end{array}\right.
\label{alfavbd}
\end{equation}
respectively. These fits are shown as solid lines in fig. \ref{fig3}.
The fitting parameters are $a_1=0.60$, $a_2=0.33$, $c_1=0.42$ and
$c_2=0.76$. Note that the different values of these parameters reflect its
dependence on the dimensionality. From eq. (\ref{alfavbd}), the Poissonian
tail of $P_c(s)$, eq. (\ref{tailsh}), is recovered for large spacings at the
limit of very strong coupling $b^d \to 0$. So, we conclude that in the case
of very strongly coupled Hamiltonians only, eq. (\ref{tailar}) losses its
validity  and eq. (\ref{tailsh}) applies.

The observed $b^d$-dependence of $\alpha$ in eq. (\ref{alfavbd})
is not surprising, since other critical properties, such as the spectral
compressibility $\chi$, or the correlation dimension $d_2$ in 1D, for which
analytical treatment is feasible, present a similar behavior towards $b$.
Specifically, $\chi=1/2\pi b$ ($b \gg 1$), $\chi=1-4b$ ($b \ll 1$),
$d_2=1-1/\pi b$ ($b \gg 1$), and $d_2=2b$  ($b \ll 1$) were derived in
refs. \cite{MF96,Mi00,ME00}. We stress that eq. (\ref{alfavbd}) is based
on numerical results and at present it should be considered as a conjecture.
So, further analytical work is needed to check this form of the critical
exponent and its origin from the model (\ref{h1dor}).

\begin{figure}[t]
\onefigure[width=7cm,height=7cm,angle=0]{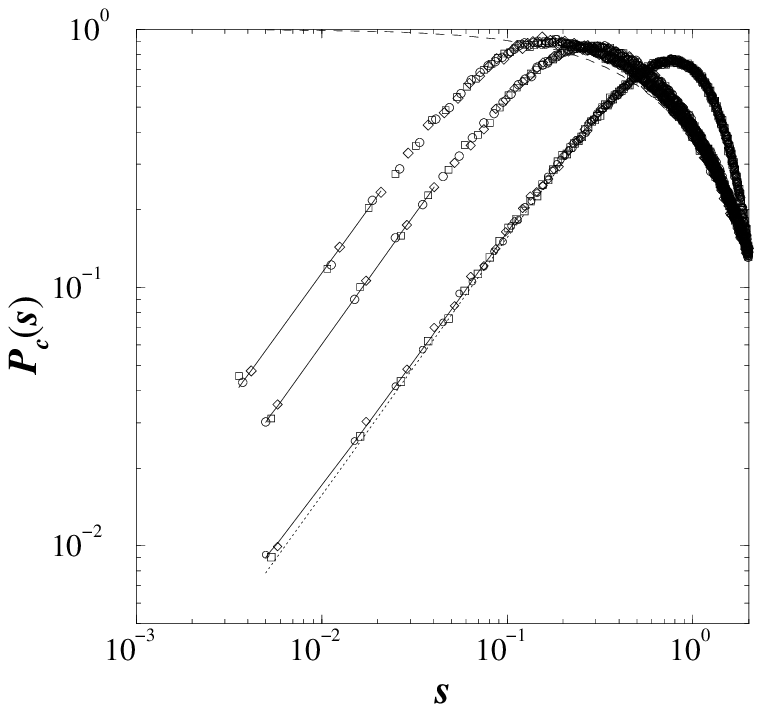}
\caption{Log-log plot of $P_c(s)$ of the 1D system for different sizes $L=1000$
(circles), 2000 (squares) and 4000 (diamonds) and disorder strength $b=5$, 0.1
and 0.05 (from bottom to top). Dotted and dashed lines are the Wigner surmise, eq.
(\ref{wigner}), and the Poisson distribution, eq. (\ref{poisson}), respectively.
Solid lines are fits, in the given intervals, to the form $P_c(s)=Cs$.}
\label{fig4}
\end{figure}

Finally, we present the limiting behavior of $P_c(s)$ as $s \to 0$.
From general considerations for the orthogonal symmetry $P_c(s) \sim s$
at small $s \ll 1$ \cite{SS93,Me91}. The results for the 1D case at
different values of $b$ for various system sizes, $L=1000$ (circles),
2000 (squares) and 4000 (diamonds), are plotted in fig. 4. Dotted and
dashed lines are the Wigner surmise, eq. (\ref{wigner}), and the Poisson
distribution, eq. (\ref{poisson}), respectively. Here we find that
$P_c (s)\sim s$ for all disorder regimes in accordance with the predictions
of refs. \cite{SS93,Me91}. The slopes of straight lines fitting the data in
the intervals shown are 1.63, 6.07, and 11.30 at $b=5$, 0.1 and 0.05,
respectively. We have checked that, for the 2D case, the same linear behavior
of $P_c(s)$ towards $s$ is fulfilled in the whole range considered of the
parameter $b^2$.

To summarize, we have investigated the critical level spacing distribution
$P_c(s)$ of non-interacting electrons on a $d$-dimensional disordered system
with long-range transfer terms in the whole range of the coupling constant
$b^{-d}$. $P_c(s)$ is found to be scale independent at all values of $b^{-d}$.
The large $s$ part of $P_c(s)$ obtained is shown to have an
$\exp [-A_ds^{\alpha}]$ decay with $1 \le \alpha \le 2$. We determined the
disorder dependence of $\alpha$ in both the strong ($b^d \ll 1$) and the weak
($b^d \gg 1$) coupling regimes. At the limit of very strong disorder $b^d \to 0$,
we found that $\alpha \to 1$ and so we obtain the expected results of the
Poissonian decay predicted in refs. \cite{SS93,AZ88}. The small-$s$ behavior
of $P_c(s) \sim s$ is in agreement with the analytical predictions at all values
of $b^d$.

\acknowledgments

\noindent
The author thanks A. M. Somoza for critically reading the manuscript
and the Spanish DGESIC for financial support through projects number
BFM2000-1059 and BFM2003-03800.

\end{document}